\documentclass[twocolumn]{revtex4}

\usepackage{amsmath}
\usepackage{amssymb}

\newtheorem{theorem}{Theorem}[section]
\newtheorem{lemma}[theorem]{Lemma.}

\newcommand{\bra}[1]{\left\langle #1 \right|}
\newcommand{\ket}[1]{\left| #1 \right\rangle}
\newcommand{\ketbra}[2]{\left| #1 \rangle \langle #2 \right|}
\newcommand{\scal}[2]{\langle #1 | #2 \rangle}

\newcommand{\opel}[3]{\left\langle #1 \left| #2 \right| #3 \right\rangle}

\begin{document}

\title{Quantum scattering theory on graphs with tails}

\author{Martin Varbanov}
\email{varbanov@usc.edu}
\affiliation{Department of Physics and Astronomy, University of Southern California, \\
Los Angeles, CA  90089}

\author{Todd A. Brun}
\email{tbrun@usc.edu}
\affiliation{Communication Sciences Institute, University of Southern California, \\
Los Angeles, CA  90089}

\date{\today}

\begin{abstract}
We consider quantum walks on a finite graphs to which infinite tails are attached. We explore how the propagating and bound states depend on the structure of the finite graph. The S-matrix for such graphs is defined. Its unitarity is proved as well as some other of its properties such as its transformation under time reversal. A spectral decomposition of the identity for the Hamiltonian of the graph is derived using its eigenvectors. We derive formulas for the S-matrix of a graph under certain operation such as cutting a tail, attaching a tail or connecting two tails to form an edge.
\end{abstract}

\maketitle

\pagenumbering{arabic}

\section{Introduction}

Quantum walks have become important tools for modeling and analyzing the behavior of a quantum system in quantum information theory. Two seemingly different types of quantum walks have been defined --- discrete-time and continuous-time quantum walks. Discrete-time quantum walks come in several flavors --- on a regular undirected graph equipped with a coin space in \cite{JWa1}, on undirected graphs in \cite{AnAl1,ViKe1}, on the edges of an undirected graph in \cite{HiBeFe1,FeHi1} or corresponding to a classical Markov chain in \cite{Sz1}. Continuous-time quantum walks first appeared in \cite{FaGu1}. Despite the similarities in their behavior, a way to obtain one as a limit of the other, something easily done in the classical case, was not known. This problem was addressed in \cite{Ch2}. It was also established that continuous-time quantum walks on sparse, low-degree graph are universal for quantum computation \cite{Ch1}.

Quantum walks, similarly to random walks in the classical case, are very useful for developing quantum algorithms. Some of the first algorithms discovered employing this approach are element distinctness \cite{Amb05}, matrix product verification \cite{BuSp1}, triangle finding \cite{MSS05} and group commutativity testing \cite{MN05}. The quantum algorithm for the glued-tree graph is based on a continuous-time quantum walk and was proven to be exponentially faster than its classical counterpart in \cite{ChCletal1}. These examples of quantum algorithms are based on the fact that the quantum walks ``hits'' a special vertex polynomially or exponentially faster when compared with the classical walk or the fastest known classical algorithm for the problem in question. Thus the question for a proper definition of a ``hitting time'' for quantum walks arises. In the discrete-time case definition for the hitting time were given and explored in \cite{JuKe1,Kr&Br1,Kr&Br2}, and for the continuous-time case in \cite{Va&Kr&Br1}. An optimal quantum algorithm for evaluating a balanced NAND tree proposed in \cite{Fa&Go&Gu1} is another example of a continuous-time quantum algorithm for which the value of the NAND tree depends on whether a certain part of the graph is hit or not. The premise for the graph is somewhat different though --- it involves the consideration of a graph with infinite tails attached to it. This result spurred a plethora of quantum algorithms for evaluating formulas \cite{AmChetal1,ReSp1}. Scattering theory for discrete-time quantum walks was developed in \cite{FeHi3,FeHi2}. Another type of quantum walk-based algorithm involves generating a sample from the uniform distribution over a graph \cite{AhAmetal1,AmBaetal1,NaVi1}. Thus one needs to define ``mixing time'' for quantum walks. Properties of mixing times and lower bounds are proven in \cite{Ric1,Ri2,KeSa1}.

In this paper we study the model considered in \cite{Fa&Go&Gu1}. There, two tails of infinite length are connected to a finite graph that contains information about the values of the variables that enter the NAND formula. A wave with fixed energy (or at least with narrowly peaked energy spectrum) is sent along a tail. After a time $T$ equal to square root of the number of variable $N$ the walk is measured to be found on one of the two tails as the wave is either totally reflected or transmitted for the selected energy. As there is a direct correspondence between the value of the reflection coefficient and the value of the NAND formula we deduce the value of the formula.

As we already noted, \cite{Ch1} proves that continuous-time quantum walks on graphs with infinite tails are universal for quantum computation. In this model, as in the previous one, the energy of the waves incoming along the tails is fixed. An arbitrary state of a qubit is represented by superposition of waves incoming along two tails. Thus $N$ qubits are represented by $2^N$ tails. A one-qubit (two-qubit) gate is a graph that has 2 (4) incoming and 2 (4) outgoing tails. Universality follows from the existance of graphs that implement the control-not, phase and Hadamard gates.

The above two papers and numerous other algorithms based on this model prompt us to explore scattering theory on graphs in continuous time. In section \ref{EnEiSt}, we give the basic model for the graph with infinite tails on which the quantum walk evolves and describe both propagating and the two kinds of bound states for the form of Hamiltonian we consider. In section \ref{OrthPB}, we prove the orthogonality between propagating and the bound states. The S-matrix is defined and its unitarity proven in section \ref{SMa}. In the following section \ref{Basis} we explore the relationship between energy eigenstates and vertex states. In sections \ref{CutTail},\ref{AttTail} and \ref{ConnTails} formulas for the S-matrix under the operations of cutting, attaching and connecting tails are derived. In the last section \ref{UnitPres} the unitarity of the S-matrix under these operations is proven.

\section{Energy eigenstates} \label{EnEiSt}

We want to introduce scattering on an undirected graph in continuous time. With each vertex $v$ of a graph $G$ we associate a normalized state $\ket{v}$ in a Hilbert space $\mathcal{H}_G$, the graph Hilbert space, with the property that states corresponding to different vertices are orthogonal to each other: $\scal{v_i}{v_j} = \delta_{ij}$. The structure of the graph determines the unitary evolution in the graph Hilbert space. The Hamiltonian we consider is a Hermitian operator on the graph Hilbert space which in the basis of vertex states is given by minus the adjacency matrix of the graph \cite{Fa&Go&Gu1}.

In order to investigate the scattering properties of graphs we need to connect ``tails'' to a finite graph. A tail is a semi-infinite linear graph with its end connected to one vertex of the graph. More than one tail can be attached to any one vertex of the graph. The vertices of each tail are number from 1 to infinity and the vertex to which it is attached is labeled with a 0. To identify different tails sometimes we will label them with the vertex of the finite graph to which they are attached. If more than one tail is attached to the same vertex we can use a second index to differentiate between different tails. The Hamiltonian on a tail is defined as in the above as minus the adjacency matrix of the tail. Explicitly
\begin{align}
H^{tail} = - \sum_{n=1}^{\infty} (\ketbra{n}{n+1} + \ketbra{n+1}{n}).\label{EQ9}
\end{align}
Let's consider a linear graph infinite in both directions. It's easy to see that the energy eigenstates $\ket{k}_{\pm}$ are given by waves propagating in the positive or negative direction: $\scal{n}{k}_{\pm} = e^{\pm i k n}$ with $k \in (0, \pi)$. States that correspond to $k = 0$ or $k = \pi$ do not behave exactly like propagating states and need to be analyzed separately. The energy corresponding to a state with absolute value of its momentum given by $|k|$ regardless of propagation direction is $E_{\pm k} = - 2 \cos k$.

\subsection{Propagating states}\label{PrSt}

We want to find the continuous energy spectrum and propagating energy states of a graph with tails. We start by considering the usual stationary Schr\"{o}dinger equation:
\begin{align}
H_{\tilde{G}} \ket{\psi} = E \ket{\psi}.\label{EQ5}
\end{align}
The Hamiltonian is given by
\begin{align}
H_{\tilde{G}} = H_G + \sum_v \sum_{m=1}^{m_v} \left( H^{tail}_{vm} - \ketbra{0_v}{1_{vm}} - \ketbra{1_{vm}}{0_v}\right).\label{EQ13}
\end{align}
In the above equation the first sum is over all vertices of $H_G$ to which tails are connected and the second sum is over all tails connected to a certain vertex $v$. The number of tails connected to vertex $v$ is denoted by $m_v$ and total number of tails by $\tilde{m} = \sum_{v} m_v$. The notation $\ket{n_{vm}}$ refers to the $n$-th vertex of the $m$-th tail connected to the vertex $v$ of the graph $G$ and as noted above $\ket{0_v}$ is the state associated to the vertex $v$, in other words $\ket{0_v} = \ket{v}$.

Let's denote by $\ket{k,vm}$, with $k \in (0, \pi)$, the propagating energy eigenstate of the above Hamiltonian which has an incoming component only on the $m$-th tail connected to the vertex $v$ of the graph $G$. We denote the restriction of this state on the graph $G$ by $\ket{k,vm}^G$. Then
\begin{align}
\scal{n_{vm}}{k,vm} = e^{- i k n} + r_{vm} (k) e^{i k n}.\label{EQ14}
\end{align}
This state on all of the rest of the tails will look like:
\begin{align}
\scal{n_{v'm'}}{k,vm} = t_{v'm',vm} (k) e^{i k n}.\label{EQ15}
\end{align}
Explicitly
\begin{align}
\ket{k,vm} = & \ket{k,vm}^G + \sum_{n=1}^{\infty} (e^{- i k n} + r_{vm} (k) e^{i k n}) \ket{n_{vm}}\notag\\
& + \sideset{}{'}\sum_{(v',m')} t_{v'm',vm} (k) \sum_{n=1}^{\infty} e^{i k n} \ket{n_{v'm'}}.\label{EQ12}
\end{align}
The prime on the sum means that we sum over all ordered pairs $(v',m')$ different from $(v,m)$, a convention we will use further in the paper. The state $\ket{k,vm}^G$ denotes the restriction of $\ket{k,vm}$ on the finite graph $G$. We note that these states are not normalizable as propagating states should be. Substituting the above expression for them in \eqref{EQ5} and taking into account that the energy corresponding to this state is $E_{k} = - 2 \cos k$, after some cancelations we obtain:
\begin{align}
H_G & \ket{k,vm}^G + (1 + r_{vm}) \ket{1_{vm}} - (e^{-i k} + r_{vm} e^{i k}) \ket{0_{v}}\notag\\
& + \sideset{}{'}\sum_{(v',m')} t_{v'm',vm} (\ket{1_{v'm'}} - e^{i k} \ket{0_{v'}})\notag\\
& - \sum_{(v',m')} \scal{0_v'}{k,vm}^G \ket{1_{v'm'}} = - 2 \cos {k} \ket{k,vm}^G. \label{EQ6}
\end{align}
As the terms that live on the tails should cancel it follows that
\begin{align}
r_{vm} + 1 & = \scal{0_{v}}{k,vm}^G,\notag\\
t_{v'm',vm} & = \scal{0_{v'}}{k,vm}^G.\label{EQ8}
\end{align}
Substituting these back in equation \eqref{EQ6} it becomes
\begin{align}
& H_G \ket{k,vm}^G - (e^{-i k} - e^{i k}) \ket{0_{v}} - e^{i k} \ket{0_{v}} \scal{0_{v}}{k}_{vm}^G\notag\\
& \, - e^{i k} \sideset{}{'}\sum_{(v',m')} \ket{0_v'} \scal{0_v'}{k,vm}^G = - 2 \cos k \ket{k,vm}^G.
\end{align}
Finally the equation that $\ket{k,vm}^G$ satisfies is
\begin{align}
\left(H_G + 2 \cos k - e^{i k} \sum_{v'} m_{v'} \ketbra{0_{v'}}{0_{v'}}\right) \ket{k,vm}^G\notag\\
= (e^{-i k} - e^{i k}) \ket{0_{v}}.
\end{align}
Another, more compact form of the equation that will turn out to be useful, we can get if we use the variable $z = e^{i k}$. Then
\begin{align}
\left(I + z H_G + z^2 Q\right) \ket{z,vm}^G = (1 - z^2) \ket{0_{v}},\label{EQ7}
\end{align}
where we introduce the following operators
\begin{align}
R & = \sum_{v'} m_{v'} \ketbra{0_{v'}}{0_{v'}},\\
Q & = I - R.
\end{align}
For convenience we denote $A(z) = I + z H_G + z^2 Q$. Later we will prove that when $z$ lies on the unit circle in the complex plane equation \eqref{EQ7} will always have a nonzero solution, and thus a propagating state will exist with reflection and transmission coefficients defined by \eqref{EQ8}. Here we want to note that when $k \in (-\pi, 0)$ the states $\ket{k,vm} = \ket{-|k|,vm}$ are defined exactly as above. They can be expressed as linear combinations of the states $\ket{|k|,vm}$. The formula for this will be given in Chapter \ref{Basis}. Another thing we want to note is that the operator $A(z)$ and the solution to \eqref{EQ7} associated with it is an instance of the so-called quadratic eigenvalue problem. For a good overview of the subject with many applications and examples discussed one can look at \cite{TiMe1}. The equation for the bound states derived in the next chapter is exactly the equation for the the right eigenvectors of $A(z)$.

The two cases $k = 0$ and $k = \pi$ need to be analyzed separately. The states corresponding to those values of $k$ still need to satisfy equation \eqref{EQ7} for $z=1$ and $z=-1$, respectively. Thus
\begin{align}
\left(I \pm H_G + Q\right) \ket{\pm 1}_{vm}^G = 0.
\end{align}
This equation may and may not have solutions, which is in contrast with the other propagating states which always exist as we shall prove below. If a solution $\ket{\epsilon}_{vm}$ with $\epsilon = \pm 1$ exists and $Q \ket{\epsilon}_{vm}^G \neq 0$ then such a state will be nonzero on some of the tails and thus it will not be square-summable. A difference from the rest of the propagating states is that there may not be linearly independent solutions for each tail. The notion of incoming and outgoing waves is also lost when it comes to these states. Thus reflection and transmission coefficients cannot be defined properly. In some respects these states behave more like bound states in the sense that they satisfy an equation identical to the bound state equation \eqref{EQ28}.

In the case when $Q \ket{\epsilon}_{vm}^G = 0$ the full state $\ket{\epsilon}_{vm}$ will be zero on each tail and thus be a real bound state of the second kind defined below.

\subsection{Bound states}

Bound states are solutions to \eqref{EQ5} which are normalizable. This is possible if and only if the amplitudes of such bound state on the tails of the graph are either exponentially decaying or zero. Thus two kinds of bound states exist. The first are states for which there is at least one tail on which the state is non-zero. The second kind of bound states are zero on all tails of the graph. We will see that the energy of these two kinds of bound states is qualitatively different.

\subsubsection{Bound states of the first kind}

Let us denote by $\ket{\varkappa_b}$ a solution of \eqref{EQ5} normalized such that on the tails of the graph it has the form
\begin{align}
\scal{n_{vm}}{\varkappa_b} \sim \alpha_{vm} (\varkappa_b) e^{- \varkappa_b n}.\label{EQ39}
\end{align}
As we want this to be a bound state of the first kind, $\alpha_{vm} \neq 0$ for at least one pair $(v,m)$. We have the following expression for the bound state
\begin{align}
\ket{\varkappa_b} = N_{\varkappa_b} \left( \ket{\varkappa_b}^G + \sum_v \sum_{m=1}^{m_v} \alpha_{vm} (\varkappa_b) \sum_{n=1}^{\infty} e^{- \varkappa_b n} \ket{n_{vm}}\right)
\end{align}
with $N_{\varkappa_b}$ being a normalization factor. Here again $\ket{\varkappa_b}^G$ is the part of the state living on the finite graph $G$. By applying the tail Hamiltonian \eqref{EQ9} to the tail of this state for which $\alpha_{vm} \neq 0$ we find its energy $E_{\varkappa_b} = - (e^{\varkappa_b} + e^{- \varkappa_b})$. For the energy to be a real number $e^{- \varkappa_b}$ should be either a real number or a complex number that belongs to the unit circle. From the requirement that the state is normalized
\begin{align}
1 = |N_{\varkappa_b}|^2 & \left({}^G\scal{\varkappa_b}{\varkappa_b}^G + \sum_v \sum_{m=1}^{m_v} \left|\alpha_{vm}\right|^2 \sum_{n=1}^{\infty} \left|e^{- \varkappa_b n}\right|^2\right)\label{EQ20}
\end{align}
it follows that the infinite sum $\sum_{n=1}^{\infty} \left|e^{- \varkappa_b n}\right|^2$ is convergent which leads to the condition that $\left|e^{- \varkappa_b}\right|^2 < 1$. This means that for bound states of the first kind $z_b = e^{- \varkappa_b}$ should be a real number with absolute value strictly less than 1 or in other words $\Re(\varkappa_b) > 0$ and $\Im(\varkappa_b) = 0$ or $\pi$.

Inserting $\ket{\varkappa_b}$ in \eqref{EQ5} leads to the following equations:
\begin{align*}
& \left(H_G + 2 \cosh \varkappa_b - e^{- \varkappa_b} R\right) \ket{\varkappa_b}^G = 0,\\
& \alpha_{vm} (\varkappa_b) = \scal{0_{vm}}{\varkappa_b}^G.
\end{align*}
If we again use a change of variables, $z_b = e^{- \varkappa_b}$, the equations take the following form:
\begin{align}
& \left(I + z_b H_G + z_b^2 Q\right) \ket{z_b}^G = 0,\label{EQ28}\\
& \alpha_{vm} (z_b) = \scal{0_{vm}}{z_b}^G.\label{EQ21}
\end{align}
We note that the operator $A(z)$ defining the propagating states through \eqref{EQ7} appears in the definition of the bound state of the first kind in the above equation but with $z = z_b$ real and satisfying $|z_b|^2 < 1$.

To determine the normalization factor $N_{z_b}$ we consider equation \eqref{EQ20} in the new variable taking into account \eqref{EQ21}. After some simplifications we obtain
\begin{align}
|N_{z_b}|^2 = \frac{1 - z_b^2}{{}^G\opel{z_b}{I-z_b^2 Q}{z_b}^G}.
\end{align}

\subsubsection{Bound states of the second kind}

In contrast with the bound states of the first kind the bound states of the second kind have zero overlap with any vertex on any tail. They live entirely on the graph G. From this it easily follows that such a bound state $\ket{\beta}$ has the property $\scal{v}{\beta} = 0$ for any vertex $v \in G$ to which a tail is attached. Because of that equation \eqref{EQ5} for a such bound state reduces to
\begin{align}
H_{G} \ket{\beta} = E_b \ket{\beta}.\label{EQ11}
\end{align}
We see that such a state is an energy eigenstate of the finite graph $G$ with energy $E_{\beta}$. We will show in the next section that if such bound states exist and their energy is less than 2 this will lead to non-invertibility of the operator $I + z H_G + z^2 Q$ for some $z$ on the unit circle. This will necessitate the redefinition of propagating states for the value of $z$.

\section{Orthogonality of propagating and bound states} \label{OrthPB}

In this section we will prove the existence of propagating states for any $k \in (0, \pi)$. The bound states of the first kind are obviously orthogonal to any propagating state because their energy, $E_{\varkappa_b} = - 2 \cosh \varkappa_b$, is always greater in absolute value than 2 and the energy of the propagating states, $E_{k} = - 2 \cos k$, is always less than or equal  in absolute value to 2 (it is well known that eigenstates of a Hermitian operator with different eigenvalues are orthogonal to each other).

Let's assume now that the operator $A (z) = I + z H_G + z^2 Q$ is not invertible for some $z = z_0$ on the unit circle. Then there is a normalized state $\ket{u}$ living totally on the finite graph such that
\begin{align}
(I + z_0 H_G + z_0^2 Q) \ket{u} = 0.\label{EQ10}
\end{align}
Multiplying the above equation on the left with $\bra{u}$ and denoting $\opel{u}{H_G}{u} = h \in \mathbb{R}$ and $\opel{u}{Q}{u} = 1 - \opel{u}{\left(\sum_v m_v \ketbra{0_v}{0_v}\right)}{u} = 1 - r \in \mathbb{R}$ we ge the following equation for $z_0$
\begin{align}
1 + z_0 h + z_0^2 (1 - r) = 0.
\end{align}
Considering the above as an equation with respect to $z_0$ its solution could be a complex number on the unit circle only if $r = \opel{u}{\left(\sum_v m_v \ketbra{0_v}{0_v}\right)}{u} = 0$. As the operator $R = \sum_v m_v \ketbra{0_v}{0_v} = \sum_v m_v P_v$ is positive, as well as a sum of positive operators, it follows that $P_v \ket{u} = \ket{0_v}\scal{0_v}{u} = 0$ for all $v \in G$ to which tails are attached. From this it is obvious that the state $\ket{u}$ is a bound state of the second kind. Equation \eqref{EQ10} reduces to
\begin{align}
((1 + z_0^2) I + z_0 H_G) \ket{u} = 0
\end{align}
which can be rewritten as
\begin{align}
H_G \ket{u} = - \left( z_0 + \frac{1}{z_0} \right) \ket{u}.
\end{align}
Comparing this to \eqref{EQ11} we find the energy of the state $\ket{u}$ to be $E_u = - (z_0 + 1/z_0)$.

We want to prove that even when $A(z_0)$ is not invertible as in the case described above, \eqref{EQ7} still has a well-defined non-zero solution which in addition is orthogonal to all bound states. First we prove that if $\ket{u}$ satisfies equation \eqref{EQ10} then it satisfies $A(z_0^*) \ket{u} = 0$ as well:
\begin{align}
A(z_0^*) \ket{u} & = (I + z_0^* H_G + (z_0^*)^2 (I - R)) \ket{u}\notag\\
& = (1 + z_0^* E_u + (z_0^*)^2) \ket{u}\notag\\
& = \left(1 - z_0^* \left(z_0 + \frac{1}{z_0}\right) + (z_0^*)^2\right) \ket{u} = 0.
\end{align}
The last equality is true because $z_0$ lies on the unit circle.

Let's denote by $K_{z_0}$ the orthonormal projector on the kernel of the operator $A (z_0)$ in the Hilbert space of $G$. In other words $K_{z_0}$ is the projector on the linear subspace spanned by all $\ket{u}$ which satisfy \eqref{EQ10}. Thus $K_{z_0}$ has the properties $A(z_0) K_{z_0} = 0$ and as we saw $K_{z_0} P_v = P_v K_{z_0} = 0$. From the above proof we also see that $A(z_0^*) K_{z_0} = 0$. Then
\begin{align}
K_{z_0} A(z_0) = (A^{\dagger} (z_0) K_{z_0}^{\dagger})^{\dagger} = (A (z_0^*) K_{z_0})^{\dagger} = 0.
\end{align}
From this it follows that $I - K_{z_0}$ is the orthonormal projector on the image of $A(z_0)$ and as $(I - K_{z_0}) P_{v'} = P_{v'} (I - K_{z_0}) = P_{v'}$ we see that $\ket{0_{v'}}$ is in the image of $A(z_0)$. Thus a solution to \eqref{EQ7} exists. To define the propagating state for $z = z_0$ we choose the only such solution that is orthogonal to kernel of $A(z_0)$ which ensures its orthogonality to all bound states of the second kind. To do this we consider the pseudo-inverse of $A (z_0)$ defined by
\begin{align}
A^{-1} (z_0) = (A (z_0) + K_{z_0})^{-1} - K_{z_0}
\end{align}
which we can use to define a solution to \eqref{EQ7} with the necessary properties:
\begin{align}
\ket{z_0,vm} & = (1 - z_0^2) A^{-1} (z_0) \ket{0_v}\notag\\
& = (1 - z_0^2) (A (z_0) + K_{z_0})^{-1} \ket{0_v}.
\end{align}
This concludes the proof of the orthogonality of propagating states and bound states.

\section{Properties of the S-matrix} \label{SMa}

The S-matrix is defined as
\begin{align}
&s_{vm,vm} = r_{vm},\\
&s_{vm,v'm'} = t_{vm,v'm'}
\end{align}
where $r_{vm}$ and $t_{vm,v'm'}$ are given by \eqref{EQ8}. The dimension of the S-matrix is equal to the number of tails attached to the graph $G$. To simplify the notation instead of using double indices for the S-matrix elements we will use just a single one, $\tau$, which will stand for the ordered pair $(v, m)$. We will also use the notation $\ket{\tau}$ for the state to which the tail labeled by $\tau = (v,m)$ is connected: $\ket{\tau} = \ket{0_v} = \ket{v}$. From \eqref{EQ7} it follows that the elements of the S-matrix will be given by
\begin{align}
s_{\tau \tau'} (z) = (1 - z^2) \opel{\tau}{A(z)^{-1}}{\tau'} - \delta_{\tau \tau'}. \label{EQ1}
\end{align}
When $A(z)$ is uninvertible, $A(z)^{-1}$ should be thought of as a pseudoinverse in the sense of the previous section.
To prove the unitarity of the S-matrix we need to introduce the following quantity
\begin{align}
j_{v_1}^{v_2} (\ket{\psi},\ket{\varphi}) = i \left(\scal{\psi}{v_2}\scal{v_1}{\varphi} - \scal{\psi}{v_1}\scal{v_2}{\varphi}\right)
\end{align}
where $\ket{\psi}, \ket{\varphi} \in \mathcal{H}_G$ and the two vertices, $v_1, v_2 \in G$, are connected to each other by an edge. A obvious property of $j_{v_1}^{v_2}$ that will be used later is the antisymmetry in the vertices: $j_{v_1}^{v_2} = - j_{v_2}^{v_1}$. Let's consider the Schr\"{o}dinger equation for the graph with the Hamiltonian given by \eqref{EQ13}
\begin{align}
i \frac{d\ket{\psi}}{dt} = H_{\tilde{G}} \ket{\psi}.
\end{align}
Any two of its solutions, $\ket{\psi(t)}, \ket{\phi(t)}$, will satisfy the following ``conservation'' equation
\begin{align}
\frac{d\scal{\psi}{v}\scal{v}{\phi}}{dt} + \sum_{v'} j_{v}^{v'} (\ket{\psi},\ket{\phi}) = 0
\end{align}
where the sum is over all vertices connected to $v$. If we define $\psi_v = \scal{v}{\psi}$ and take $\ket{\psi(t)} = \ket{\phi(t)}$ the equation takes the more standard form
\begin{align}
\frac{d\psi_v^* \psi_v}{dt} + \sum_{v'} j_{v}^{v'} (\ket{\psi},\ket{\psi}) = 0.
\end{align}
The quantity $\psi_v^* \psi_v$ is just the probability at vertex $v$ at time $t$ and thus $j_{v}^{v'} (\ket{\psi},\ket{\psi})$ can be thought of as the probability current at time $t$ flowing from vertex $v$ to vertex $v'$.

Let's assume now that $\ket{\psi(0)}$ and $\ket{\varphi(0)}$ are eigenstates of $H_{\tilde{G}}$ with equal energy $E$. Then
\begin{align}
\frac{d\psi_v^* \varphi_v}{dt}  = 0
\end{align}
and so
\begin{align}
\sum_{v'} j_{v}^{v'} (\ket{\psi},\ket{\varphi}) = 0
\end{align}
for any vertex $v \in \tilde{G}$. Summing the formula above over all vertices in $G$ we have
\begin{align}
\sum_{v \in G} \sum_{v'} j_{v}^{v'} (\ket{\psi},\ket{\varphi}) = 0.
\end{align}
In the above sum for each term $j_{v}^{v'} (\ket{\psi},\ket{\varphi})$ there exist a term $j_{v'}^{v} (\ket{\psi},\ket{\varphi})$ whenever $v,v' \in G$. Because of the antisymmetry of $j_{v}^{v'}$ every such pair of two term will cancel out, leaving us with terms for which one of the vertices is on a tail:
\begin{align}
\sum_{v \in G} \sum_{v'} j_{v}^{v'} (\ket{\psi},\ket{\varphi}) = \sum_{\tau'} j_{0_{\tau'}}^{1_{\tau'}}(\ket{\psi},\ket{\varphi}) = 0.
\end{align}
If we choose $\ket{\psi} = \ket{\varphi} = \ket{z,\tau}$ (the solution to equation \eqref{EQ5} which when restricted to $\mathcal{H}_G$ satisfied equation \eqref{EQ7}) the above equation reduces to
\begin{align}
|s_{\tau \tau}|^2 + \sideset{}{'}\sum_{\tau'} |s_{\tau' \tau}|^2 = 1.
\end{align}
Here we use the convention the a primed sum signifies that we sum over all $\tau' \neq \tau$. If we choose $\ket{\psi} = \ket{z,\tau_1}$ and $\ket{\varphi} = \ket{z,\tau_2}$ we obtain
\begin{align}
\sum_{\tau'} s_{\tau' \tau_1}^* s_{\tau' \tau_2} = 0.
\end{align}
This proves the unitarity of the S-matrix.

Another property that we will need is
\begin{align}
s_{\tau \tau'}(z^*) = (s^{\dagger})_{\tau \tau'}(z) = s^*_{\tau' \tau}(z).\label{EQ36}
\end{align}
Equation \eqref{EQ36} is easy to prove using \eqref{EQ1}:
\begin{align*}
s_{\tau \tau'}(z^*) & = (1 - (z^*)^2) \opel{\tau}{A(z^*)^{-1}}{\tau'} - \delta_{\tau \tau'}\\
& = (1 - z^2)^* \opel{\tau}{(A(z)^{\dagger})^{-1}}{\tau'} - \delta_{\tau \tau'}\\
& = (1 - z^2)^* \opel{\tau'}{A(z)^{-1}}{\tau}^* - \delta_{\tau \tau'}\\
& = \left((1 - z^2) \opel{\tau'}{A(z)^{-1}}{\tau} - \delta_{\tau \tau'}\right)^*\\
& = s_{\tau' \tau}^*(z) = (s^{\dagger})_{\tau \tau'}(z).
\end{align*}

\section{Vertex state basis and energy eigenvalue basis}\label{Basis}

In this section we present formulas for the spectral decomposition of the identity corresponding to the Hamiltonian $H_{\tilde{G}}$. It is easy to obtain after determining its eigenstates --- the propagating states together with the bound states of the first and second kind. For $v, v' \in \tilde{G}$ we have
\begin{align}
\delta_{v v'} = \sum_{b} \psi_{b}(v) & \psi_{b}^*(v') + \sum_{\beta} \psi_{\beta}(v) \psi_{\beta}^*(v')\notag\\
& + \frac{1}{2 \pi} \int\limits_{0}^{\pi} \sum_{\tau} \psi_{k,\tau}(v) \psi_{k,\tau}^*(v') dk,
\end{align}
where
\begin{align*}
& \psi_{b}(v) = \scal{v}{\varkappa_b},\\
& \psi_{\beta}(v) = \scal{v}{\beta},\\
& \psi_{k,\tau}(v) = \scal{v}{k,\tau}.
\end{align*}
In Dirac notation the formula will give the resolution of the identity operator $I_{\mathcal{H}_{\tilde{G}}}$ for the Hilbert space $\mathcal{H}_{\tilde{G}}$:
\begin{align}
I_{\mathcal{H}_{\tilde{G}}} = \sum_{b} & \ketbra{\varkappa_b}{\varkappa_b} + \sum_{\beta} \ketbra{\beta}{\beta}\notag\\
&\;\;\;\; + \frac{1}{2 \pi} \int\limits_{0}^{\pi} \sum_{\tau} \ket{k,\tau} \bra{k,\tau} dk,
\end{align}
Using this formula we can easily obtain a simple expression for a vertex state $\ket{n_{\tau}}$ corresponding to a vertex on the tail $\tau$ of the graph $\tilde{G}$ in terms of the energy eigenstates:
\begin{align}
\ket{n_{\tau}} = \sum_{b} & \scal{\varkappa_b}{n_{\tau}} \ket{\varkappa_b} + \sum_{\beta} \scal{\beta}{n_{\tau}} \ket{\beta}\notag\\
& + \frac{1}{2 \pi} \int\limits_{0}^{\pi} \sum_{\tau'} \scal{k,\tau'}{n_{\tau}} \ket{k,\tau'} dk.\label{EQ38}
\end{align}
This can be further simplified. As we have already proven, the bound states of the second kind have zero overlap with any vertex that lies on a tail, $\scal{\beta}{n_{\tau}}=0$. Thus the second sum in the above expression is zero. Formula \eqref{EQ39} also implies that $\scal{\varkappa_b}{n_{\tau}} = e^{-\varkappa_b n} \scal{\varkappa_b}{0_{\tau}}$. From \eqref{EQ14} and \eqref{EQ15} we see that
\begin{align}
\scal{k,\tau'}{n_{\tau}} = \delta_{\tau \tau'} e^{ikn} + s_{\tau \tau'}^*(k) e^{-ikn},\label{EQ35}
\end{align}
or in terms of the $z$ variable
\begin{align}
\scal{z,\tau'}{n_{\tau}} = \delta_{\tau \tau'} z^n + s_{\tau \tau'}^*(z) z^{-n},\label{EQ37}
\end{align}
Here we need the following identity:
\begin{align}
\sum_{\nu} s_{\nu \tau} (k) \ket{-k,\nu} = \ket{k,\tau},\label{EQ31}
\end{align}
or in terms of the $z$ variable
\begin{align}
\sum_{\nu} s_{\nu \tau} (z) \ket{z^*,\nu} = \ket{z,\tau}.\label{EQ34}
\end{align}
First we prove this for the restriction of the propagating states on the graph $G$:
\begin{widetext}
\begin{align}
\sum_{\nu} \ket{z^*,\nu}^{G} s_{\nu \tau} (z) & = \sum_{\nu} (1 - {z^*}^2) A^{-1} (z^*) \ket{\nu} \left((1 - z^2)\opel{\nu}{A^{-1} (z)}{\tau} - \delta_{\nu \tau}\right)\notag\\
& = \frac{(z^2 - 1)}{z^2} A^{-1 \dagger} (z) \left((1 - z^2) R A^{-1} (z) - I\right) \ket{\tau}\notag\\
& = (1 - z^2) A^{-1 \dagger} (z) \left(z^{* 2} A(z) - z^{* 2} (1 - z^2) R\right) A^{-1} (z) \ket{\tau}\notag\\
& = (1 - z^2) A^{-1 \dagger} (z) \left(z^{* 2} I + z^* H + I - R - z^{* 2} R +R\right) A^{-1} (z) \ket{\tau}\notag\\
& = (1 - z^2) A^{-1 \dagger} (z) A^{\dagger} (z) A^{-1} (z) \ket{\tau} = \ket{z,\tau}^{G}
\end{align}
\end{widetext}
In the above proof we used that $|z|=1$. All inverses above are well defined as well because they act on $\ket{\tau}$ which lies in the image of $A (z)$.

Now we need to prove the identity for any vertex lying on a tail. Multiplying \eqref{EQ34} by $\ket{n_{\tau'}}$ we arrive at the following formula which we need to prove:
\begin{align}
\sum_{\nu} s_{\nu \tau} (z) \scal{n_{\tau'}}{z^*,\nu} = \scal{n_{\tau'}}{z,\tau}.\notag
\end{align}
It is easily proven using \eqref{EQ37} and \eqref{EQ36}:
\begin{align}
\sum_{\nu} & s_{\nu \tau} (z) \scal{n_{\tau'}}{z^*,\nu} = \sum_{\nu} s_{\nu \tau} (z) \scal{z^*,\nu}{n_{\tau'}}^*\notag\\
& = \sum_{\nu} s_{\nu \tau} (z) (\delta_{\tau' \nu} z^n + s_{\tau' \nu}(z^*) z^{-n})\notag\\
& = \sum_{\nu} s_{\nu \tau} (z) (\delta_{\tau' \nu} z^n + (s^{\dagger})_{\tau' \nu}(z) z^{-n})\notag\\
& = s_{\tau' \tau} (z) z^n + \delta_{\tau' \tau} z^{-n} = (\delta_{\tau' \tau} z^n + s_{\tau' \tau}^* (z) z^{-n})^*\notag\\
& = \scal{z,\tau}{n_{\tau'}}^* = \scal{n_{\tau'}}{z,\tau}.
\end{align}
This proves \eqref{EQ31}.

Now we can simplify the sum under the integral in \eqref{EQ38} with the help of \eqref{EQ35},\eqref{EQ31} and the unitarity of the S-matrix:
\begin{align*}
& \sum_{\tau'} \scal{k,\tau'}{n_{\tau}} \ket{k,\tau'}\\
& = \sum_{\tau'} (\delta_{\tau \tau'} e^{ikn} + s_{\tau \tau'}^*(k) e^{-ikn}) \ket{k,\tau'}\\
& = e^{ikn} \ket{k,\tau} + e^{-ikn} \ket{-k,\tau}.
\end{align*}
Substituting this back into \eqref{EQ38} leads to:
\begin{align}
\ket{n_{\tau}} & = \sum_{b} \scal{\varkappa_b}{n_{\tau}} \ket{\varkappa_b}\notag\\
&\;\;\; + \frac{1}{2 \pi} \int\limits_{0}^{\pi} (e^{ikn} \ket{k,\tau} + e^{-ikn} \ket{-k,\tau}) dk\notag\\
& = \sum_{b} \scal{z_b}{n_{\tau}} \ket{z_b} + \frac{1}{2 \pi} \int\limits_{-\pi}^{\pi} e^{ikn} \ket{k,\tau} dk.\label{EQ40}
\end{align}
Using the $z$ variable this formula takes very appealing form:
\begin{align}
\ket{n_{\tau}} & = \sum_{b} z_b^n \scal{z_b}{0_{\tau}} \ket{z_b} + \frac{1}{2 \pi i} \oint\limits_{C} z^n \ket{z,\tau} \frac{dz}{z}.
\end{align}

\section{Cutting a Tail}\label{CutTail}

In this section we want to investigate the effect of cutting a tail on the $S$-matrix of the graph. We will show that the $S$-matrix of the new, pruned graph can be expressed in terms of the elements of the $S$-matrix of the old graph. Thus we are given a graph $G$ to which $n$ tails are attached and we assume that we know its $S$-matrix. Without loss of generality we can choose to cut the first tail which we denote by $\tau_c$. For the unpruned graph and pruned graph the elements of the $S$-matrix are given by \eqref{EQ1} with
\begin{align}
A(z) & = I + z H_G + z^2 Q,\label{EQ2}\\
A_c(z) & = I + z H_G + z^2 Q_c = A + z^2 P_{\tau_c}\notag\\
& = A + z^2 \ketbra{\tau_c}{\tau_c},
\end{align}
respectively. As can be seen from \eqref{EQ1} the problem reduces to expressing the matrix elements of $A_c^{-1}$ in terms of the matrix elements of $A^{-1}$.
\begin{align}
& \bra{\tau} A_c^{-1} \ket{\tau'}\notag\\
& = \bra{\tau} (A + z^2 P_{\tau_c})^{-1} \ket{\tau'}\notag\\
& = \bra{\tau} (I + z^2 A^{-1} P_{\tau_c})^{-1} A^{-1} \ket{\tau'}\notag\\
& = \bra{\tau} \sum_{j=0}^{\infty} (- z^2 A^{-1} P_{\tau_c})^j A^{-1} \ket{\tau'}\notag\\
& = \bra{\tau} A^{-1} \ket{\tau'} - z^2 \bra{\tau} A^{-1} \ket{\tau_c}\notag\\
&\;\;\;\; \times \sum_{j=1}^{\infty} (- z^2 \bra{\tau_c} A^{-1} \ket{\tau_c})^{j-1} \bra{\tau_c} A^{-1} \ket{\tau'}\notag\\
& = \bra{\tau} A^{-1} \ket{\tau'} - \frac{z^2 \bra{\tau} A^{-1} \ket{\tau_c} \bra{\tau_c} A^{-1} \ket{\tau'}}{1 + z^2 \bra{\tau_c} A^{-1} \ket{\tau_c}}.\notag
\end{align}
Multiplying both sides of this expression by $1 - z^2$ and using \eqref{EQ1} we get
\begin{align}
s^c_{\tau \tau'} = s_{\tau \tau'} - \frac{z^2 s_{\tau \tau_c} s_{\tau_c \tau'}}{1 + z^2 s_{\tau_c \tau_c}}.\label{EQ16}
\end{align}

\subsection{Leaving a Stump}

In stead of cutting the tail $\tau_c$ at the root now we want to leave a stump of length $L$. This case can be easily subsumed in the previous one. We define a new tail $\tilde{\tau}_c$ which coincides with $\tau_c$ but its beginning is at the $L$-th vertex of the tail $\tau_c$, $\ket{L_{\tau_c}}$. Thus $\ket{0_{\tilde{\tau}_c}} = \ket{L_{\tau_c}}$ The propagating solution to \eqref{EQ5} in this case will be equal up to a phase to the solution when we think of the tail being attached to the root $\ket{0_{\tau_c}}$. The conditions that the reflection and transmission coefficients need to satisfy is given by \eqref{EQ14} and \eqref{EQ15} which is only possible when the phase between the two solutions is chosen appropriately: $\ket{k}_{\tilde{\tau}_c} = e^{i k L} \ket{k}_{\tau_c}$. Thus
\begin{align}
& \scal{n_{\tilde{\tau}_c}}{k}_{\tilde{\tau}_c} = e^{i k L} \scal{(n+L)_{\tau_c}}{k}_{\tau_c}\notag\\
& = e^{- i k n} + e^{2ikL} r_{\tau_c} e^{ikn} = z^{-n} + z^{2L} r_{\tau_c} z^n,\notag\\
& \scal{n_{\tau}}{k}_{\tilde{\tau}_c} = e^{ikL} \scal{n_{\tau}}{k}_{\tau_c} = e^{ikL} t_{\tau \tau_c} e^{ikn} = z^L t_{\tau \tau_c} z^n.\notag
\end{align}
For energy eigenstates with the same energy but which are incoming on a different tail we get
\begin{align}
& \scal{n_{\tilde{\tau}_c}}{k}_{\tau} = \scal{(n+L)_{\tau_c}}{k}_{\tau} = e^{ikL} t_{\tau_c \tau} e^{ikn} = z^L t_{\tau \tau_c} z^n.\notag
\end{align}
All other such conditions that don't involve the tail that is being cut are satisfied automatically. From the above equations and the definitions of the reflection and transmission coefficients \eqref{EQ8} we see that the elements of the $S$-matrix for the graph with tail $\tilde{\tau}_c$ satisfy:
\begin{align}
\tilde{s}_{\tilde{\tau}_c \tilde{\tau}_c} & = z^{2 L} s_{\tau_c \tau_c},\notag\\
\tilde{s}_{\tilde{\tau}_c \tau} & = z^L s_{\tau_c \tau},\notag\\
\tilde{s}_{\tau \tilde{\tau}_c} & = z^L s_{\tau \tau_c},\notag\\
\tilde{s}_{\tau \tau'} & = s_{\tau \tau'}.\notag
\end{align}
Plugging those in \eqref{EQ16} we finally get:
\begin{align}
s^c_{\tau \tau'} = s_{\tau \tau'} - \frac{z^{2(L + 1)} s_{\tau \tau_c} s_{\tau_c \tau'}}{1 + z^{2(L + 1)} s_{\tau_c \tau_c}}.
\end{align}

\subsection{Cutting $k$ Tails}

It is obvious that if we cut more than one tail it doesn't matter the order in which we cut them. Let us assume that the $S$-matrix has the following block form:
\begin{align}
S = \left(
\begin{array}{cc}
T_{(\tilde{m}-k) \times (\tilde{m}-k)} & U_{(\tilde{m}-k) \times k}\\
V_{k \times (\tilde{m}-k)} & W_{k \times k}\\
\end{array}
\right).\label{EQ18}
\end{align}
Here $\tilde{m}$, as defined in Chapter \ref{EnEiSt}, is the total number of tails and $k$ is the number of tails that will be cut. Thus we want to cut the tails corresponding to the last $k$ entries in the $S$-matrix. We obtain the following formula for the new $S$-matrix:
\begin{align}
S^c = T - z^2 U (I + z^2 W)^{-1} V.\label{EQ17}
\end{align}
(The identity matrix in the above formula is a $k \times k$ matrix.) It is obvious that formula \eqref{EQ16} follows from the above.

\section{Attaching a Tail}\label{AttTail}

Let's assume that we want to attach one more tail to a vertex that already has at least one tail attached to it. We want to express the $S$-matrix of the new graph in term of the old one. Of course, it is also possible to add a tail to a vertex that doesn't already have one, but in this case the new $S$-matrix cannot be derived simply from the $S$-matrix of the original graph. Let us denote the new tail being attached with $\tau_a$, the vertex we attach it to with $\tilde{v}$ and the tail already attached to the vertex $\tilde{v}$ with $\tau_{\tilde{v}}$. We note that $\ket{\tau_{\tilde{v}}} = \ket{\tau_a} = \ket{\tilde{v}}$. By similar argument to the one we made in section \ref{CutTail} we get
\begin{align}
\bra{\tau} A_a^{-1} \ket{\tau'} = \bra{\tau} A^{-1} \ket{\tau'} + \frac{z^2 \bra{\tau} A^{-1} \ket{\tau_c} \bra{\tau_c} A^{-1} \ket{\tau'}}{1 - z^2 \bra{\tau_c} A^{-1} \ket{\tau_c}},\notag
\end{align}
where
\begin{align}
A_a = A - z^2 P_{\tau_c} = A - z^2 \ketbra{\tau_c}{\tau_c}.\notag
\end{align}
In order to reexpress everything in terms of the elements of the old $S$-matrix we consider different cases depending on the first and second indices of the $S$-matrix. In the below $\tau$ and $\tau'$ signify tails that are different from both $\tau_a$ and ${\tilde{\tau}}$. After some simplification we get
\begin{align}
s^a_{\tau \tau'} & = s_{\tau \tau'} + \frac{z^2 s_{\tau \tau_{\tilde{v}}} s_{\tau_{\tilde{v}} \tau'}}{1 - z^2(2 + s_{\tau_{\tilde{v}} \tau_{\tilde{v}}})},\notag\\
s^a_{\tau \tau_{\tilde{v}}} & = s^a_{\tau \tau_a} = \frac{(1 - z^2) s_{\tau \tau_{\tilde{v}}}}{1 - z^2(2 + s_{\tau_{\tilde{v}} \tau_{\tilde{v}}})},\notag\\
s^a_{\tau_{\tilde{v}} \tau} & = s^a_{\tau_a \tau} = \frac{(1 - z^2) s_{ \tau_{\tilde{v}} \tau}}{1 - z^2(2 + s_{\tau_{\tilde{v}} \tau_{\tilde{v}}})},\\
s^a_{\tau_{\tilde{v}} \tau_a} & = s^a_{\tau_a \tau_{\tilde{v}}} = \frac{(1 - z^2) (1 + s_{\tau_{\tilde{v}} \tau_{\tilde{v}}}) }{1 - z^2(2 + s_{\tau_{\tilde{v}} \tau_{\tilde{v}}})},\notag\\
s^a_{\tau_{\tilde{v}} \tau_{\tilde{v}}} & = s^a_{\tau_a \tau_a} = \frac{z^2 + s_{\tau_{\tilde{v}} \tau_{\tilde{v}}}}{1 - z^2(2 + s_{\tau_{\tilde{v}} \tau_{\tilde{v}}})}.\notag
\end{align}

\section{Connecting two tails to form an edge}\label{ConnTails}

The setup is as in the previous sections but the goal now is to connect two tails --- in other words to cut two of the tails connected to the graph and replace them with an edge between the vertices they were connected to. Again without loss of generality we connect the first and second tails, $\tau_1$ and $\tau_2$. When we do that the Hamiltonian and the operator $Q$ change to
\begin{align*}
\tilde{H}_G & = H_G - X_{(2)},\\
\tilde{Q} & = Q + P_{(2)}
\end{align*}
where we have used the notation
\begin{align}
X_{(2)} & = \ketbra{\tau_1}{\tau_2} + \ketbra{\tau_2}{\tau_1},\notag\\
P_{(2)} & = \ketbra{\tau_1}{\tau_1} + \ketbra{\tau_2}{\tau_2}.\notag
\end{align}
This leads to the following expression for $\tilde{A}(z)$
\begin{align}
\tilde{A} & = I + z \tilde{H}_G + z^2 \tilde{Q} = A - z X_{(2)} + z^2 P_{(2)}.\notag
\end{align}
For convenience we denote $B =z^2 P_{(2)} - z X_{(2)}$. Again we look at the matrix elements of $\tilde{A}^{-1}$:
\begin{align}
& \bra{\tau} \tilde{A}^{-1} \ket{\tau'} = \bra{\tau} (A + B)^{-1} \ket{\tau'}\notag\\
& = \bra{\tau} (I + A^{-1} B)^{-1} A^{-1} \ket{\tau'} = \bra{\tau} \sum_{j=0}^{\infty} (- A^{-1} B)^j A^{-1} \ket{\tau'}\notag
\end{align}
From the definition of $B$ it follows that
\begin{align}
P_{(2)} B P_{(2)} = P_{(2)} B = B P_{(2)} = B,
\end{align}
from which we see that
\begin{align}
& \bra{\tau} \tilde{A}^{-1} \ket{\tau'} - \bra{\tau} A^{-1} \ket{\tau'}\notag\\
& = \sum_{j=1}^{\infty} \bra{\tau} (- A^{-1} P_{(2)} B P_{(2)})^j A^{-1} \ket{\tau'}\notag\\
& = - \bra{\tau} A^{-1} P_{(2)} \sum_{j=1}^{\infty} B (- P_{(2)} A^{-1} P_{(2)} B)^{j-1} P_{(2)} A^{-1} \ket{\tau'}\notag\\
& = - \bra{\tau} A^{-1} P_{(2)} B (P_{(2)} + P_{(2)} A^{-1} P_{(2)} B)^{-1} P_{(2)} A^{-1} \ket{\tau'}\notag\\
& = - \bra{\tau} A^{-1} P_{(2)} (B^{-1} + P_{(2)} A^{-1} P_{(2)})^{-1} P_{(2)} A^{-1} \ket{\tau'}.\label{EQ4}
\end{align}
In the above whenever necessary the inverses should be thought of as pseudo-inverses.

We want to use a block-matrix representation again. We will use an almost identical notation as the one we used before
\begin{align}
S = \left(
\begin{array}{cc}
T_{n \times n} & U_{n \times 2}\\
V_{2 \times n} & W_{2 \times 2}\\
\end{array}
\right),
\end{align}
but now the last two entries in the $S$-matrix correspond to the tails that are to be connected. Multiplying \eqref{EQ4} by $1 - z^2$ and after some simplifications we get:
\begin{align}
\tilde{S} = T - z U (z W - X)^{-1} V\label{EQ19}
\end{align}
with $X$ being just the $X$ Pauli matrix.

To give an explicit formula for the elements of the new $S$-matrix we need the pseudo-inverses of $B^{-1} + P_{(2)} A^{-1} P_{(2)}$:
\begin{align}
& \left(B^{-1} + P_{(2)} A^{-1} P_{(2)}\right)^{-1}\notag\\
& = \frac{1}{D(z)} \left(B - z^2 (1 - z^2) C_{(2)}\right)\label{EQ3}
\end{align}
where
\begin{align}
D(z) & = 1 - (a_{12} + a_{21}) z + (a_{11} + a_{22}) z^2\notag\\
& \, - (a_{11} a_{22} - a_{12} a_{21}) z^2 (1 - z^2),\notag\\
C_{(2)} & = a_{22} \ketbra{\tau_1}{\tau_1} - a_{12} \ketbra{\tau_1}{\tau_2}\notag\\
& - a_{21} \ketbra{\tau_2}{\tau_1} + a_{11} \ketbra{\tau_2}{\tau_2},\notag
\end{align}
where $a_{ij}$ are defined through the following formula:
\begin{align}
P_{(2)} A^{-1} P_{(2)} & = a_{11} \ketbra{\tau_1}{\tau_1} + a_{12} \ketbra{\tau_1}{\tau_2}\notag\\
& + a_{21} \ketbra{\tau_2}{\tau_1} + a_{22} \ketbra{\tau_2}{\tau_2}.
\end{align}
Substituting \eqref{EQ3} in \eqref{EQ4} and again using the definition of the elements of the S-matrix \eqref{EQ1} we find
\begin{align}
& \tilde{s}_{\tau \tau'} = s_{\tau \tau'}\notag\\
& + \frac{z (s_{\tau \tau_1} s_{\tau_2 \tau'} + s_{\tau \tau_2} s_{\tau_1 \tau'})}{1 - (s_{\tau_1 \tau_2} + s_{\tau_2 \tau_1}) z - (s_{\tau_1 \tau_1} s_{\tau_2 \tau_2} - s_{\tau_1 \tau_2} s_{\tau_2 \tau_1}) z^2}\notag\\
& - \frac{z^2 (s_{\tau_1 \tau_2} s_{\tau \tau_1} s_{\tau_2 \tau'} + s_{\tau_2 \tau_1} s_{\tau \tau_2} s_{\tau_1 \tau'})}{1 - (s_{\tau_1 \tau_2} + s_{\tau_2 \tau_1}) z - (s_{\tau_1 \tau_1} s_{\tau_2 \tau_2} - s_{\tau_1 \tau_2} s_{\tau_2 \tau_1}) z^2}\notag\\
& + \frac{z^2 (s_{\tau_2 \tau_2} s_{\tau \tau_1} s_{\tau_1 \tau'} + s_{\tau_1 \tau_1} s_{\tau \tau_2} s_{\tau_2} \tau')}{1 - (s_{\tau_1 \tau_2} + s_{\tau_2 \tau_1}) z - (s_{\tau_1 \tau_1} s_{\tau_2 \tau_2} - s_{\tau_1 \tau_2} s_{\tau_2 \tau_1}) z^2}.
\end{align}

\subsection{Composition of unitary gates}

The universality of quantum walks in continuous time was proved in \cite{Ch1}. In the model presented there a quantum wire corresponds to a set of tails connected to a graph. Specifically the state of a qudit is represented by a linear superposition of incoming waves on $d$ of the tails with the same energy. Thus to each tail corresponds one of the orthogonal states of the qudit. The graph to which the tails are connected will implement the quantum gate. Another set of $d$ tails connected to the graph represent the quantum wire which carries the state of the qudit with the quantum gate applied to it. The order for this to represent a quantum gate it is needed that a wave coming in on any incoming tail needs to scatter on outgoing tails only for the fixed energy at which the computation is performed (the scattering matrix of a graph cannot satisfy this condition for every energy). Thus it is easy to see that for that particular energy the S-matrix for this graph needs to have a block form:
\begin{align}
S=\left(
\begin{array}{cc}
0 & S_{io}\\
S_{oi} & 0\\
\end{array}
\right)
\end{align}
where the first $d$ entries stand for incoming tails and the last $d$ for outgoing. Thus the quantum gate being implemented is given by the unitary matrix $S_{oi}$.

We want to show that if we connect two graph using the rules for connecting tails we will end up with a gate that is composition of the quantum gates corresponding to each graph. Before we have connected the two graphs the S-matrix is just the direct product of the S-matrices:
\begin{align}
S = \left(
\begin{array}{cccc}
0 & S_{io}^1 & 0 & 0\\
S_{oi}^1 & 0 & 0 & 0\\
0 & 0 & 0 & S_{io}^2\\
0 & 0 & S_{oi}^2 & 0
\end{array}
\right).\label{EQ41}
\end{align}
We will use a generalization of formula \eqref{EQ19} to find the S-matrix after the connections are being made. The formula retains the same form even if we connect more than two tails as long as we arrange the tails being connected to correspond to entries in the lower right corner of the S-matrix. The generalization $\tilde{X}$ of $X$ in formula \eqref{EQ19} is going to be given by either
\begin{align}
\tilde{X} = \left(
\begin{array}{cccc}
X & 0 & 0 & 0\\
0 & X & 0 & 0\\
0 & 0 & \ddots & 0\\
0 & 0 & 0 & X
\end{array}
\right)
\end{align}
or
\begin{align}
\tilde{X} = \left(
\begin{array}{cc}
0 & I\\
I & 0\\
\end{array}
\right)
\end{align}
depending on how the tails being connected are arranged in the matrix $W$. Thus by permuting the entries \eqref{EQ41} we obtain:
\begin{align}
S' = \left(
\begin{array}{cccc}
0 & 0 & 0 & S_{io}^1\\
0 & 0 & S_{oi}^2 & 0\\
0 & S_{io}^2 & 0 & 0\\
S_{oi}^1 & 0 & 0 & 0
\end{array}
\right).
\end{align}
Thus for the block matrices $T, U, V$ and $W$ we have
\begin{align}
T = 0, \;\;\;U = \left(
\begin{array}{cc}
0 & S_{io}^1\\
S_{oi}^2 & 0
\end{array}
\right),\\
V = \left(
\begin{array}{cc}
0 & S_{io}^2\\
S_{oi}^1 & 0
\end{array}
\right), \;\;\;W = 0
\end{align}
and $\tilde{X}$ needs to be of the second form. For the new S-matrix we obtain:
\begin{align}
\tilde{S} & = T - z U (z W - X)^{-1} V\\
& = - z \left(
\begin{array}{cc}
0 & S_{io}^1\\
S_{oi}^2 & 0
\end{array}
\right)
\left(
\begin{array}{cc}
0 & -I\\
-I & 0
\end{array}
\right)^{-1}
\left(
\begin{array}{cc}
0 & S_{io}^2\\
S_{oi}^1 & 0
\end{array}
\right)\\
& = z \left(
\begin{array}{cc}
0 & S_{io}^1 S_{io}^2\\
S_{oi}^2 S_{oi}^1 & 0
\end{array}
\right)
\end{align}
We see that that unitary being implemented is given by $z S_{oi}^2 S_{oi}^1$ which is exactly equal to the composition of the two unitaries up to the phase $z$. The edges that connect the two graphs which are formed after connecting the tails lead to the introductions of this phase.

\section{Unitarity preservation in cutting and connecting tails}\label{UnitPres}

In this section we prove that the above operations preserve the unitary of the S-matrix. We need the following lemma.
\begin{lemma}
Given a unitary matrix $S$ with block-form representation
\begin{align}
S = \left(
\begin{array}{cc}
T_{n \times n} & U_{n \times m}\\
V_{m \times n} & W_{m \times m}\\
\end{array}
\right),\notag
\end{align}
and an unitary matrix $G_{m \times m}$ such that $G+W$ is invertible, the matrix
\begin{align}
\tilde{S} = T - U (G + W)^{-1} V
\end{align}
is unitary.
\end{lemma}
{\it Proof.} The unitarity of $S$ implies that
\begin{align}
T T^{\dagger} & = I - U U^{\dagger}, & T^{\dagger} T & = I - V^{\dagger} V,\notag\\
T^{\dagger} U & = - V^{\dagger} W, & U^{\dagger} T & = - W^{\dagger} V,\notag\\
T V^{\dagger} & = - U W^{\dagger}, & V T^{\dagger} & = - W U^{\dagger},\notag\\
V V^{\dagger} & = I - W W^{\dagger}, & U^{\dagger} U & = I - W^{\dagger} W.
\end{align}
In the above formulas, although not indicated, the identity $I$ needs to be understood as having the appropriate dimension for each equation. Then
\begin{widetext}
\begin{align}
\tilde{S} \tilde{S}^{\dagger} & = \left(T - U (G + W)^{-1} V\right) \left(T - U (G + W)^{-1} V\right)^{\dagger}\notag\\
& =  T T^{\dagger} - U (G + W)^{-1} V T^{\dagger} - T V^{\dagger} (G^{\dagger} + W^{\dagger})^{-1} U^{\dagger} + U (G + W)^{-1} V V^{\dagger} (G^{\dagger} + W^{\dagger})^{-1} U^{\dagger}\notag\\
& = I - U U^{\dagger} + U (G + W)^{-1} W U^{\dagger} + U W^{\dagger} (G^{\dagger} + W^{\dagger})^{-1} U^{\dagger} + U (G + W)^{-1} (I - W W^{\dagger}) (G^{\dagger} + W^{\dagger})^{-1} U^{\dagger}\notag\\
& = I - U \left(I - (G + W)^{-1} W - W^{\dagger} (G^{\dagger} + W^{\dagger})^{-1} - (G + W)^{-1} (G^{\dagger} + W^{\dagger})^{-1} + (G + W)^{-1} W W^{\dagger} (G^{\dagger} + W^{\dagger})^{-1} \right) U^{\dagger}\notag\\
& = I - U \left((I - (G + W)^{-1} W) (I - W^{\dagger} (G^{\dagger} + W^{\dagger})^{-1}) - (G + W)^{-1} (G^{\dagger} + W^{\dagger})^{-1}\right) U^{\dagger}\notag\\
& = I - U \left((G + W)^{-1} G G^{\dagger} (G^{\dagger} + W^{\dagger})^{-1} - (G + W)^{-1} (G^{\dagger} + W^{\dagger})^{-1}\right) U^{\dagger} = I.
\end{align}
\end{widetext}
This proves the lemma.

In the case of cutting a tail from formula \eqref{EQ17} we see that $G = I/z^2$ which is unitary because $|z| = 1$.

In the case of connecting two tails from formula \eqref{EQ19} we see that $G = - X/z$ which is obviously unitary as well.

This proves the unitarity of the $S$-matrix after an operation of cutting or connecting tails.

\section{Discussion}

Quantum walks in their many varieties have proven immensely useful in the area of Quantum Computation as a tool for developing new algorithms or as an abstract model to studying the behavior of quantum systems. The model we are concerned with in this paper involves quantum walks on infinite graphs, it is natural to develop scattering theory for these systems. We have studied both propagating and bound states, and how they depend on the structure of the finite graph to which the tails are attached. The equations that define the propagating and bound states are an example of the quadratic eigenvalue problem. The S-matrix, its unitarity, definition and conservation of probability density and current were addressed, as well as the orthogonality of propagating and bound states. We were able to derive formulas for the S-matrix of a graph that is obtained by operations on the tails in terms of the S-matrix of the original graph.

In future work, we would like to pursue the definition of hitting time for such graphs. In previous work, \cite{JuKe1,Kr&Br1,Kr&Br2,Va&Kr&Br1}, properties of the hitting time for finite graphs were explored. We would like to give appropriate definition for hitting time when we consider the tailed graphs. One such definition may follow in the footsteps of \cite{Va&Kr&Br1}, where the quantum walk starts from an arbitrary state, and we are interested in the probability and average time to hit any of the tails. A different approach could be followed if we consider a wave packet that is incoming along one of the tails with an energy in a narrow band. Then a definition for a hitting time may be the time it takes for the packet to reach a different outgoing tail to some standard graph. For example, if we have a graph with just two tails, we can compare the time for which a packet travels through the graph from one of the tails to the other and compare it with the time that the same packet will take to move along the infinite line.

Defining the hitting time will permit us to explore new ideas and paradigms for quantum algorithms. In \cite{Kr&Br2} infinite hitting times were discovered for the discrete-time case, and later the same phenomenon was observed for the continuous-time case in \cite{Va&Kr&Br1}. The existence of an infinite hitting times can lead to possible quantum algorithms. The algorithms can be based on the simple observation that the quantum walk may or may not hit a certain set of vertices depending on the existence of infinite hitting time. This would serve as the final result in our quantum computation, much like the evaluation of the NAND tree is based on whether the quantum walk hits the second tail or not. Both infinite hitting times and exponentially fast quantum algorithms as in \cite{ChCletal1} are based on the idea that the quantum walk is constrained to evolve on a subspace of the whole Hilbert space because some special property of the graph. Symmetry of the graph is one property that can lead to these effects, but other properties may do so as well.  This is an open problem, and an area of active research.

\begin{acknowledgements}
We would like to thank Shesha Raghunathan for helpful conversations. This work was supported in part by NSF Grant No. EMT-0829870 and NSF Grant No. CCF-0448658.
\end{acknowledgements}

\end{document}